\documentstyle[12pt]{article}
\begin{document}
\title{Determination of polarized parton
distributions in the nucleon
\thanks{Work supported in part by the KBN-Grant 2-P302-143-06.}}
\author{Jan Bartelski\\
Institute of Theoretical Physics, Warsaw University,\\
Ho$\dot{z}$a 69, 00-681 Warsaw, Poland. \\ \\
\and
Stanis\l aw  Tatur \\
Nicolaus Copernicus Astronomical Center,\\ Polish Academy of Sciences,\\
Bartycka 18, 00-716 Warsaw, Poland.\\}
\date{}
\maketitle

\begin{abstract}
\noindent A fit to proton, neutron and deuteron spin asymmetries
is presented and polarized parton distributions in nucleon are
given. These densities have their roots in the MRS fit for
unpolarized case. The integrals of polarized distributions are
compared with the experimental figures. The role
of polarized gluons is also discussed.

\end{abstract}

\newpage
There are a lot of new experiments which yield the data on spin
structure of a nucleon. We have the newest experimental points from
CERN \cite{d,p} and SLAC \cite{n,e143,e143d} and also the older
ones from these laboratories \cite{pold}, \cite{emc}.
The spin asymmetries are measured on proton
(SLAC-Yale \cite{pold}, EMC \cite{emc}, SMC \cite{p} and
E143 \cite{e143}), neutron (E142 \cite{n} made on $^{3}He$) and
deuterium (SMC \cite{d} and E143 \cite{e143d}) targets.
Using the experimental figures we can study phenomenologically
nucleon spin structure and in
particular we can determine the polarized parton distributions.

For unpolarized case several fits were performed [8-11], there
were also attempts to get the spin distributions [12-18]. Some time
ago Martin, Roberts
and Stirling (MRS) \cite{mrs} presented a complete fit
with determination of  parton (i.e. \frenchspacing quark and gluon)
distributions. Quite recently after the measurements from Hera
\cite{hera} for the small
$x$ region the new fits with more reliable gluon contribution
and modified sea were performed \cite{mrs2}.
Our discussion how to determine the polarized
structure functions using MRS fits \cite{mrs}
was given in Ref.\cite{bt} and in Ref.\cite{bt1}.

In this paper we would like to get polarized quark parton distributions
starting from the unpolarized ones and using existing data
on spin asymmetries. The calculated
values of octet
axial-vector couplings $a_{3}$ and $a_{8}$ obtained from the fit are compared
with the experimental values gotten from nucleon and hyperon
$\beta$-decays and modified for QCD corrections.
We use these quantities to find the best fit (of course together with
the $\chi^{2}$ values). In order to stabilize the fits we put
the restriction on $a_{8}$ (such procedure takes place also e.g.
in Ref.\cite{Ger.Stir}).
We have also tried to include polarized gluons contributing in the way
proposed in Ref.{\cite{ross}. It comes out that the polarized gluonic
degrees of freedom  do\ not lead to any substantial improvement in a fit.

Let us start with the formulas for unpolarized quark parton distributions
given (at $ Q^{2}=4\, {\rm GeV^{2}}$) in the newest fit performed
by Martin, Roberts and Stirling \cite{mrs2}. We have for the
valence quarks:
\begin{eqnarray}
u_{v}(x)&=&2.704 x^{-0.407}(1-x)^{3.96}(1-0.76\sqrt{x}+4.20x),
\nonumber \\
d_{v}(x)&=&0.251 x^{-0.665}(1-x)^{4.41}(1+8.63\sqrt{x}+0.32x) ,
\end{eqnarray}

\noindent and for the antiquarks from the sea:
\begin{eqnarray}
2\bar{u} (x)&=&0.392M(x)-\delta (x), \nonumber \\
2\bar{d} (x)&=&0.392M(x)+\delta (x), \nonumber \\
2\bar{s} (x)&=&0.196M(x), \\
2\bar{c} (x)&=&0.020M(x). \nonumber
\end{eqnarray}

\noindent In eq.(2) the singlet contribution is:
\begin{equation}
M(x)=1.74 x^{-1.067}(1-x)^{10.1}(1-3.45\sqrt{x}+10.3x),
\end{equation}

\noindent whereas the isovector part:
\begin{equation}
\delta (x)=0.043x^{-0.7}(1-x)^{10.1}(1+64.9x).
\end{equation}

\noindent For the unpolarized gluon distribution we get:
\begin{equation}
G(x)=1.51x^{-1.301}(1-x)^{6.06}(1-4.14\sqrt{x}+10.1x).
\end{equation}

We assume, in an analogy to the unpolarized case, that the polarized
quark distributions are of the form:
$x^{\alpha}(1-x)^{\beta}P_{2}( \sqrt{x})$, where
$P_{2}(\sqrt{x})$ is a second order polynomial in $\sqrt{x}$ and
the asymptotic behaviour for $x$$\rightarrow$$0$ and
$x$$\rightarrow$$1$ (i.e. the values of $\alpha$ and $\beta$)
are the same (except for $\Delta M$, see a discussion below) as in
unpolarized case.
Our idea is to split the numerical constants
(coefficients of $P_{2}$ polynomial)  in
eqs.(1, 3 and 4)  in two parts in
such a manner that the distributions are positive
defined.
Our expressions for $\Delta q(x) = q^{+}(x)-q^{-}(x)$
($q(x) = q^{+}(x)+q^{-}(x)$) are:
\begin{eqnarray}
\Delta u_{v}(x)&=&x^{-0.407}(1-x)^{3.96}(a_{1}+a_{2}\sqrt
{x}+a_{3}x), \nonumber \\
\Delta
d_{v}(x)&=&x^{-0.665}(1-x)^{4.41}(b_{1}+b_{2}\sqrt{x}+b_{3}x), \nonumber \\
\Delta M(x)&=&x^{-0.567}(1-x)^{10.1}(c_{1}+c_{2}\sqrt{x}), \\
\Delta \delta (x)&=&x^{-0.7}(1-x)^{10.1}c_{3}(1+64.9x). \nonumber
\end{eqnarray}

\noindent
For a moment we will not take into account polarized gluons
i.e. we put $\Delta G =0$.
For total sea polarization i.e. $\Delta M$, we
assume that there is no
term behaving like $x^{-1.067}$ at small $x$ (we assume that $\Delta M$
and hence all sea distributions are integrable), which means that
coefficient in this case have to be splitted into equal parts in
$M^{+}$ and $M^{-}$. The next term ($x^{-0.567}$) is relatively more
singular then the sea contribution in the preferred fit in
Ref.\cite{bt1}. On the other hand when we put the coefficient in
front of $x^{-0.567}$ equal to zero (i.e. $c_{1}=0$) $M(x)$ will
behave for $x \rightarrow 0$ like $x^{-0.067}$. In this case we
get the fit with higher $\chi^{2}/N_{DF}$. That means that in
spite of the fact that in the present fit the unpolarized sea
behaviour is less singular for small $x$ values, contrary to the
case in Ref.\cite{bt1}, the model for $\Delta M(x)$ with more
singular sea contribution is phenomenologicaly chosen. As we
will see that influences the behaviour of $g_{1}^{p}(x)$ for
small $x$ values.

In order to get the unknown parameters in the expressions for polarized
quark distributions at $ Q^{2}=4\, {\rm GeV^{2}}$ (see eq.(6)) we make
a fit  to the experimental data on spin asymmetries for proton, neutron and
deuteron targets.
The theoretical expressions for these asymmetries are given by:
\begin{eqnarray}
A^{p}_{1}(x)&=&\frac{4\Delta u_{v}(x)+\Delta d_{v}(x)+2.236
\Delta M(x)-3\Delta \delta (x)}{4u_{v}(x)+d_{v}(x)+2.236
M(x)-3\delta (x)}(1+R),  \nonumber \\
A^{n}_{1}(x)&=&\frac{\Delta u_{v}(x)+4\Delta d_{v}(x)+2.236
\Delta M(x)+3\Delta \delta (x)}{u_{v}(x)+4d_{v}(x)+2.236
M(x)+3\delta (x)}(1+R), \\
A^{d}_{1}(x)&=&\frac{5\Delta u_{v}(x)+5\Delta d_{v}+4.472\Delta
M(x)} {5u_{v}(x)+5d_{v}+4.472M(x)}
(1-\frac{3}{2}p_{D})(1+R). \nonumber
\end{eqnarray}

Where the ratio $R=\sigma_{L}/\sigma_{T}$ (which vanishes in the
Bjorken limit) is taken
from \cite{whit} and $p_{D}$ is a probability of D-state in deuteron
wave function (equal to $(5\pm 1)\%$ \cite{d,e143d}.

{\em We assume} that the spin asymmetries do not depend on
$Q^{2}$ (it is only our first order approximation) what is
suggested by the experimental data \cite{d,n} and
phenomenological analysis \cite{alt}.
We hope that
numerically our results  at $ Q^{2}=4\, {\rm GeV^{2}}$ will not change
much if the evolution of $F_1$ and $g_1$ functions will be taken into
account.

Spin structure function e.g.
$g_{1}^{p}$ is given by:
\begin{equation}
g_{1}^{p}(x)=(4\Delta u_{v}(x)+\Delta d_{v}(x)+2.236
\Delta M(x)-3\Delta \delta (x))/18.
\end{equation}

The obtained polarized quark distributions $\Delta u(x)$,
$\Delta d(x)$, $\Delta M(x)$ and $\Delta\delta(x)$ can be used
to calculate first
moments. For a given $Q^{2}$ we can write the
relations:
\begin{eqnarray}
\Gamma^{p}_{1}& = &\frac{4}{18}\Delta u+\frac{1}{18}\Delta d+
\frac{1}{18}\Delta s+\frac{4}{18}\Delta c, \nonumber \\
\Gamma^{n}_{1}& = &\frac{1}{18}\Delta u+\frac{4}{18}\Delta d+
\frac{1}{18}\Delta s+\frac{4}{18}\Delta c ,
\end{eqnarray}

\noindent where $\Delta q=\int^{1}_{0}\Delta q(x)\, dx$ and
$\Gamma_{1}=\int^{1}_{0}g_{1}(x)\, dx$.

We define other combinations of integrated quark polarizations:
\begin{eqnarray}
a_{3}& = &\Delta u-\Delta d , \nonumber \\
a_{8}& = &\Delta u+\Delta d-2\Delta s , \\
\Delta\Sigma& =&\Delta u+\Delta d+\Delta s , \nonumber
\end{eqnarray}

Such results for the integrated quantities (calculated at $4\,{\rm GeV^{2}}$)
after taking into account known QCD corrections (see e.g. Ref.\cite{lar})
could be compared with axial-vector coupling constants
$g_{A}$ and $g_{8}$ known from neutron $\beta$-decay and
hyperon $\beta$-decays (in the last case one needs $SU(3)$
symmetry). The difference of $\Gamma^{p}_{1}(Q^{2})$ and
$\Gamma^{n}_{1} (Q^{2})$ can be expressed by:
\begin{equation}
\Gamma^{p}_{1}(Q^{2})-\Gamma^{n}_{1}(Q^{2})=(\Delta u(Q^2)-\Delta
d(Q^2))/6=a_{3}(Q^2)/6=c_{NS}(Q^{2})g_{A}/6,
\end{equation}

\noindent
 where $c_{NS}(Q^{2})$ describes
QCD corrections for non-singlet quantities \cite{lar}
and $g_{A}=1.2573\pm 0.0028$ (see e.g. Ref.\cite{cr})
is obtained from the neutron $\beta$-decay.
In our paper $Q^2$ is constant and takes the value $4\,{\rm GeV^{2}}$.

We get from the experimental figure $a_{3}(4\,{\rm GeV^{2}})=
c_{NS}(4\,{\rm GeV^{2}})g_{A}=1.11$
and with this value we shall compare $a_{3}$ calculated from our fits.
Another useful combination of $\Gamma^{p}_{1}(Q^{2})$ and
$\Gamma^{n}_{1}(Q^{2})$ is equal to:
\begin{equation}
\Gamma^{p}_{1}(Q^{2})+\Gamma^{n}_{1}(Q^{2})=5a_{8}(Q^{2})/18+2\Delta
s(Q^{2})/3
\end{equation}

\noindent
with $a_{8}=c_{NS}(Q^{2})g_{8}$, where $g_{8}=0.58\pm 0.03$
\cite{cr} is obtained from the hyperon $\beta$-decays.
Knowing $c_{NS}(Q^{2})$ we can calculate $a_{8}(4\,{\rm
GeV^{2}})=0.51\pm 0.03$
and with this number we shall compare the results obtained from our fit.
If we have had very precise experimental data and in the whole $x$ range
there would be
no problems with determination of polarized quark distributions. Unfortunately
that is not the case yet. Actually, from the experiment we have
information on $\Gamma^{p}_{1}$ and $\Gamma^{n}_{1}$.
The combination $\Gamma^{p}_{1}$-$\Gamma^{n}_{1}$ is directly
connected to $g_{A}$ experimental
quantity modified by QCD corrections.  On
the other hand $\Gamma^{p}_{1}$+$\Gamma^{n}_{1}$ is the
combination of $a_{8}$ and $\Delta s$ and it came out in
Ref.\cite{bt1} that
the fits are not sensitive enough to determine $a_{8}$ and $\Delta s$
separately in a stable way. The values of $a_{8}$ and $\Delta s$
were different for our models and for
different subsets of data. To stabilize the determination of
parameters we also here assume in addition that $a_{8}=0.51$ with
$0.1$ as artificial theoretical error.

We get the following values of our parameters (describing the polarized
quark distributions in eq.(6)) from the fit to all existing data for spin
asymmetries:
\begin{equation}
\begin{array}{lll}
a_{1}=\hspace*{0.293cm} 1.71,&a_{2}=-6.68,&a_{3}=\hspace*{0.101cm} 14.2,\\
b_{1}=-0.005,&b_{2}=\hspace*{0.36cm} 0.835,&b_{3}=-3.33,\\
c_{1}=-1.10,&c_{2}=\hspace*{0.35cm} 1.38,&c_{3}=-0.03.
\end{array}
\end{equation}

\marginpar{\em Table 1}
In the second row of the Table 1 the integrated quantities:
$\Gamma^{p}_{1}$, $\Gamma^{n}_{1}$, $a_{3}$, $a_{8}$,
$\Delta\Sigma$ and $\Delta M$ together with the coresponding
$\chi^{2}/N_{DF}$ that follow from our fit are presented. They
can be compared (first row) with the values obtained from our
previous fit presented in Ref.\cite{bt1}. We will not show the comparison
of the results of the fit to CERN data only (proton + deuteron)
and SLAC data (proton + neutron + deuteron). Two fits like
before are in agreement with each other and with the overall fit.
 It is specially interesting because the CERN data
are taken at much smaller $x$ values then the SLAC data.

In Figs.(1, 2 and 3) we present the comparison of our fit  with the
experimental asymmetries for proton (1), neutron (2) and deuterium (3) target.
We see that in the case of deuterium many experimental points
lie below our fitted curve especially in the large $x$ region
where the errors are relatively big. In the Fig.(2) we see that
$A_{1}^{n}$ approaches very slowly zero for $x\rightarrow 0$
what produces relatively high (negative) value of
$\Gamma_{1}^{n}$. Because we have
62 points for proton spin asymmetries in comparison with 33 for
deuteron and 8 for neutron it seems that our curves are dominated by the
proton data. Small discrepancies between different experiments
are also not excluded.

The obtained quark distributions lead
to the following integrated quantities:
$\Delta u=0.69$ ($\Delta u_{v}=0.83$), $\Delta d=-0.45$
($\Delta d_{v}=-0.08$) and $\Delta s=-0.13$,
hence the amount of sea polarization is $\Delta M=-0.65$.
This last number is not small and hence also the strange sea polarization
in our model is rather big. As one can see from Table 1 the $a_{3}$ value
seems to be in our fit very close to the predicted value 1.11.
$\Delta\Sigma$ is rather small and the tendency that the model with
more singular sea behaviour ($c_{1}\neq 0$) produces very small
$\Delta\Sigma$ values is confirmed (this was already observed in
Ref.\cite{bt1}).
Our integrated quantities ($\Gamma^{p}_{1}$ and $\Gamma^{n}_{1}$)
differ slightly from the values quoted by the experimental groups, whose
figures are calculated directly from
the experimental points with the assumption of Regge type behaviour at small
$x$. But on the other hand our polarized quark distributions satisfy all
the constraints taken implicitly into account in fits to the
unpolarized data.
The difference in integrated quantities comes mainly due to
our assumptions about small $x$ behaviour for spin-dependent distributions.
If we compare  our results integrated in the interval $0.029<x<0.8$ (region
covered in the SLAC experiments) we get $\Gamma^{p}_{1}$=0.11 (SLAC result
without the extrapolation to unmeasured region is $0.12\pm 0.01$)
and $\Gamma^{d}_{1}=0.04$ (to be compared with $0.04\pm 0.005$).
Also we use $R$ \cite{whit} which is not very realiable for small $x$ and
this also gives an additional error, which is rather difficult to estimate.

It is interesting to see what will happen when we relax our
condition $a_{8}=0.51\pm 0.1$. We get the fit with comparable
$\chi^2$ per degree of freedom ($\chi^2/N_{DF}=1.02$) and
$\Gamma_{1}^{p}=0.12$,
$\Gamma_{1}^{n}=-0.08$, $a_{3}=1.23$, $a_{8}=0.23$, $\Delta
\Sigma=0.13$ and $\Delta M=-0.18$. We see that $a_{8}$ value (
$a_{8} \sim \Delta \Sigma$ ) is
rather small in this case and also we have much smaller sea
polarization.

We have also tried to include polarized gluons along the line of
ref.\cite{ross} assuming for the gluon distribution:
\begin{equation}
\Delta G(x)=x^{-0.801}(1-x)^{6.06}(d_{1}+d_{2}\sqrt{x}),
\end{equation}

\noindent with a new $d_{1}$ and $d_{2}$ constants which have to
be fitted.
The appearance of non-zero gluonic distribution affects our formulas
only through the substitution:
$\Delta q \Rightarrow \Delta q-\frac{\alpha_{s}}{2\pi}\Delta G$.
In such fit ($\chi^{2}/N_{DF}=1.01$) we got (after integration)
the negative sign of the gluonic
contribution (if we take gluonic distribution with the full strength at
$x=0$). In this case $\Gamma_{1}^{p}$ is equal to 0.19 because the gluonic
contribution is added to the quark contribution instead of being
substracted.

In the row three of Table 1 the results of the fit with gluons
are shown in which gluon contribution is less singular $\Delta
G(x)\sim x^{-0.301}$ for small $x$  i.e. the
coefficient in front of $x^{-0.801}$ is equaly devided between
$G^{+}(x)$ and $G^{-}(x)$ ($d_{1}=0$). In this case $\chi^{2}/N_{DF}=1.02$
and is comparable to the fit without gluons. In this case sea
polarization is very small and also $\Delta\Sigma$, as expected,
is rather small ($0.06$). Also we get in this case for $a_{3}$
the value higher than the experimental figure. We conclude that
the inclusion of gluonic
contribution does not lead to the substantial improvement of the
fit.

The Fig.(4) shows the comparison of $g^{p}_{1}(x)$ calculated
from our fit with the experimental points (evolved to common
value $Q^2=4\,{\rm GeV^{2}}$). We do not observe the
growth of $g^{p}_{1}$ for small $x$ values in our model contrary
to the previous fit in Ref.\cite{bt1}. It is caused by the
relatively singular behaviour of the sea contribution for small
$x$ values.

Starting from the new, improved version of the MRS fit \cite{mrs2}
to the unpolarized deep inelastic data we have made a fit to
proton, neutron ($^{3}He$) and deuteron spin asymmetries in
order to obtain polarized quark parton distributions.
To stabilize the fits we added
the experimental information on octet quantity $a_{8}$. We have
calculated the parameters of the polarized quark distributions
using the combination of all existing proton, neutron and
deutron spin asymmetries measurements (including the new results on proton
and deutron E143 experiments and very recent SMC deuteron
data from CERN).
We do not need gluonic contributions to be taken into account, i.e. the fit
with gluons is not better.
The next step in front of us is to include $Q^{2}$ dependence of spin
asymmetries in comparison with experimental data.

\newpage
{\bf Figure captions}

\begin{itemize}
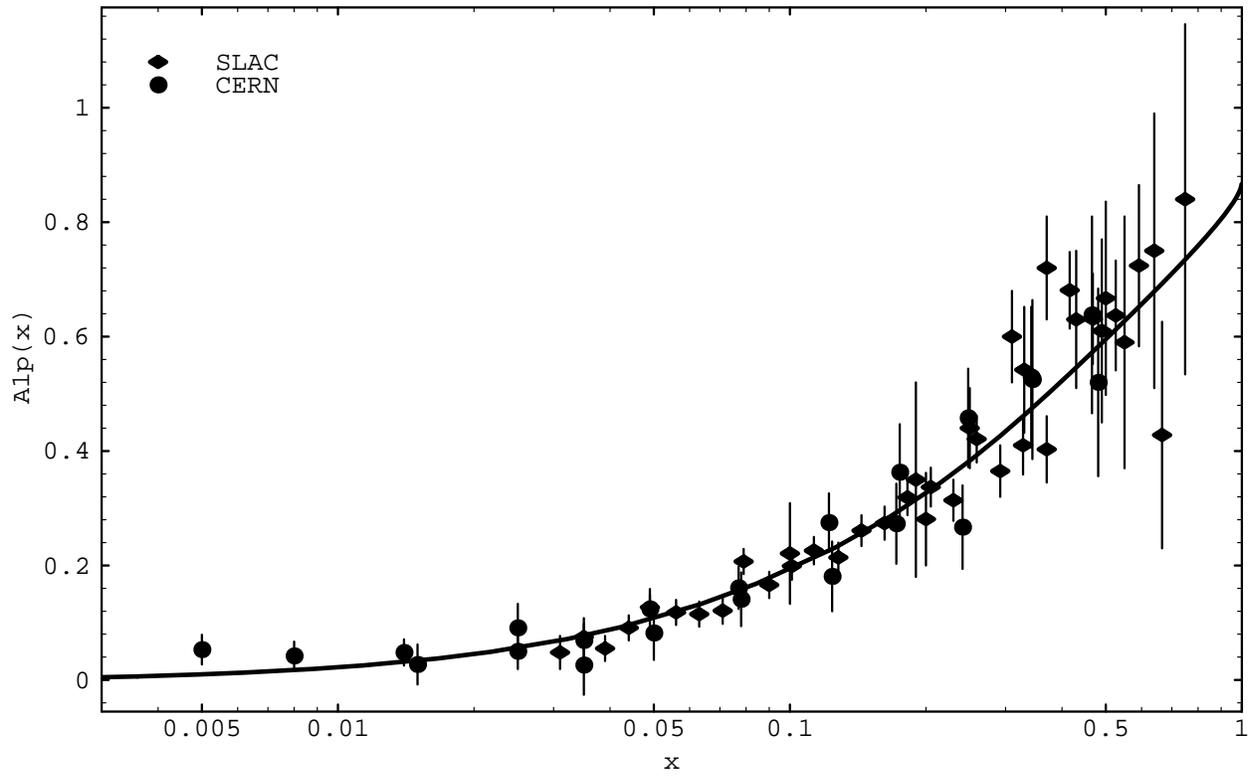
\item[ Figure 1 \ ] The comparison of spin asymmetry on protons
with the curve gotten from our fit to all existing data.
Points are taken from SLAC (E80, E130, E143) and CERN (EMC, SMC) experiments.
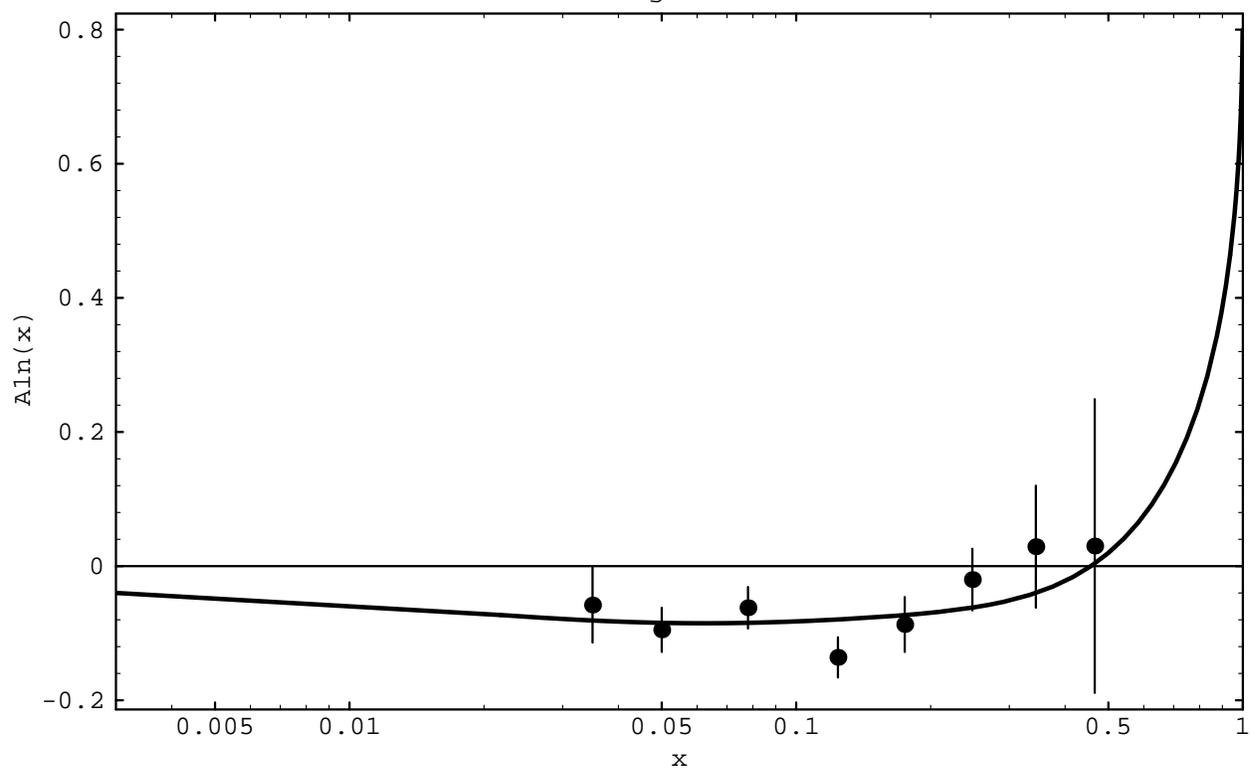
\item[Figure 2 \ ] The comparison of spin asymmetry on neutrons
(SLAC E142 data) with the curve gotten from our fit.
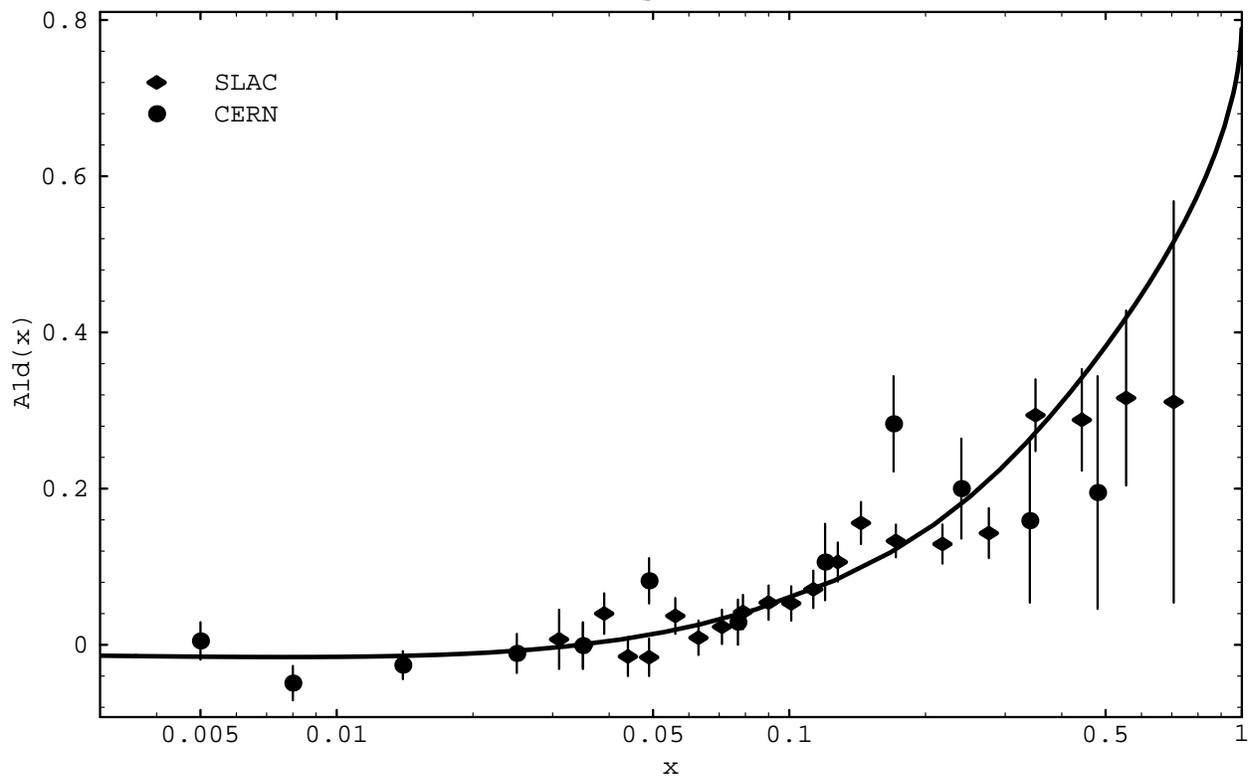
\item[Figure 3 \ ] Our prediction for deuteron asymmetry compared
with the SMC and SLAC data.
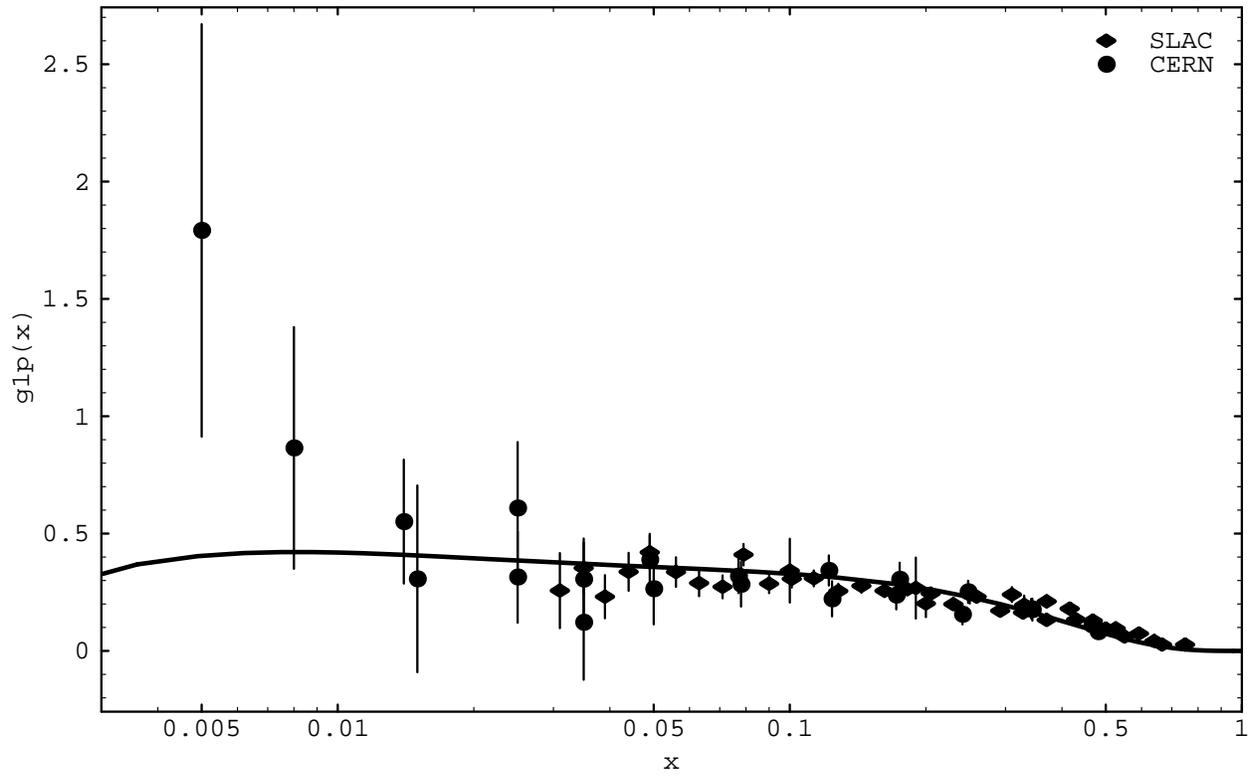
\item[Figure 4 \ ] The data for $g_{1}^{p}(x)$ structure function
with the curve gotten using the parameters of our fit. The data points
are taken from SLAC  and CERN and evolved to common
$ Q^{2}=4\, {\rm GeV^{2}}$.
\end{itemize}

\newpage

\begin{center}
{\bf Table 1}
\end{center}
 The first moments of polarized distributions (see
eqs.(9) and (10)). The strange sea polarization $\Delta s$ (not
presented in the Table) is
connected to the total sea polarization by the relation:
$\Delta s=0.196 \Delta M$.
We have made our fits taking all existing
experimental data on spin asymmetries.
\begin{center}
\begin{tabular}{|c||r|r|r|r|r|r||r|} \hline
{\em Fit}&$\Gamma^{p}_{1}$&$\Gamma^{n}_{1}$&$a_{3}$&$a_{8}$&$\Delta\Sigma$&
$\Delta M$&$\chi^{2}/N_{DF}$ \\ \hline\hline
{\em Old fit [17] }&0.139&-0.072&1.27&0.47&0.20&-0.45&1.05 \\
\hline {\em New fit}&0.119&-0.072&1.14&0.50&0.12&-0.65&1.01 \\ \hline
{\em New fit with
gluons}&0.123&-0.082&1.23&0.51&0.06&-0.09&1.02 \\  \hline
\end{tabular} \\[18pt]
\end{center}

\newpage
\begin{thebibliography}{99}
\bibitem{d} B.Adeva {\em et al.} (Spin Muon Collaboration), Phys.Lett.
{\bf B 302}, 533 (1993); D.Adams {\em et al.} (Spin Muon Collaboration),
Phys.Lett. {\bf B 357}, 248 (1995);
\bibitem{p} D.Adams {\em et al.} (Spin Muon Collaboration), Phys.Lett.
{\bf B 329}, 399 (1994);
\bibitem{n} D.L.Anthony {\em et al.} (E142 Collaboration), Phys.Rev.Lett.
{\bf 71}, 959 (1993);
\bibitem{e143} K.Abe {\em et al.} (E143 Collaboration), Phys.Rev.Lett.
{\bf 74}, 346 (1995);
\bibitem{e143d} K.Abe {\em et al.} (E143 Collaboration), Phys.Rev.Lett.
{\bf 75}, 25 (1995);
\bibitem{pold} M.J.Alguard {\em et al.} (SLAC-Yale Collaboration),
Phys.Rev.Lett.
{\bf 37}, 1261 (1976); G.Baum {\em et al.}, Phys.Rev.Lett. {\bf 45},
2000 (1980); {\bf 51}, 1135 (1983);
\bibitem{emc} J.Ashman {\em et al.} (European Muon Collaboration),
Phys.Lett. {\bf B 206}, 364 (1988); Nucl. Phys. {\bf B 328}, 1 (1989);
\bibitem{mrs} A.D.Martin, W.J.Stirling and R.G.Roberts, Phys.Rev. {\bf D 47},
867 (1993); Phys. Lett. {\bf B 306}, 145 (1993); {\em erratum} {\bf B 309},
492 (1993);
A.D.Martin, W.J.Stirling and R.G.Roberts,
Phys.Rev. {\bf D 50}, 6734 (1994);
\bibitem{mrs2} A.D.Martin, W.J.Stirling and R.G.Roberts,
Phys.Lett. {\bf B 354}, 155 (1995);
\bibitem{fits} J.F.Owens, Phys.Lett. {\bf B 266}, 126 (1991);
P.Chiapetta, G.Nardulli, Z.Phys. {\bf C 51}, 435 (1991);
L.W.Whitlow {\em et al.}, Phys.Lett. {\bf B 282}, 475, (1992);
M.Gl\"{u}ck, E.Reya, A.Vogt, Z.Phys. {\bf C 53}, 127 (1992);
M.Gl\"{u}ck, E.Reya, A.Vogt, Phys.Lett. {\bf B 306}, 391 (1993);
J.Botts {\em et al.} (CTEQ Collaboration), Phys.Lett. {\bf B
304}, 159 (1993);
\bibitem{brod} S.J.Brodsky, M.Burkardt, I.Schmidt, Nucl.Phys.
{\bf B 441}, 197 (1995);
\bibitem{Ger.Stir} T.Gehrmann, W.J.Stirling, Z.Phys. {\bf C
65}, 461 (1995);
\bibitem{Bucc} F.Buccella {\em et al.}, preprint Napoli
DSF-T-95/26 (1995);
\bibitem{Bour.Sof} C.Bourrely, J.Soffer, Nucl.Phys. {\bf B 445},
341 (1995);
\bibitem{GRV} M.Gl\"{u}ck {\em et al.} preprint Dortmund-TH
95/13 (1995);
\bibitem{bt} J.Bartelski, S.Tatur, Acta Phys. Pol. {\bf B 26},
913 (1994);
\bibitem{bt1} J.Bartelski, S.Tatur, CAMK preprint No. 288 to be
published in Z.Phys.;
\bibitem{ross} G.Altarelli, G.G.Ross, Phys.Lett. {\bf B 212}, 391 (1988);
\bibitem{hera} T.Ahmed {\em et al.}, ($H_{1}$ Collaboration),
Desy preprint 95-006 (1995); M.Derrick {\em et al.}, (ZEUS Collaboration),
Z.Phys. {\bf C 65}, 379 (1995); M.Derrick {\em et al.},
Phys.Lett. {\bf B 345}, 576 (1995);
I.Abt {\em et al.} ($H_1$ Collaboration), Phys.Lett.
{\bf B 321}, 161 (1994);
\bibitem{whit} W.L.Whitlow {\em et al.} Phys.Lett. {\bf B 250}, 193 (1990);
\bibitem{alt} G.Altarelli, P.Nason, G.Ridolfi, Phys.Lett. {\bf B 320}, 152
(1994);
\bibitem{lar} S.A.Larin, Phys.Lett. {\bf B 334}, 192 (1994);
\bibitem{cr} F.E.Close and R.G.Roberts, Phys.Lett. {\bf B 316}, 165 (1993).
\end{thebibliography}
\end{document}